\documentclass[12pt,fleqn]{article}

\usepackage{epsfig}
\usepackage{amsmath}
\usepackage{times}
\usepackage[sort&compress,comma]{natbib}
\usepackage{amssymb}

\title{\sffamily\textbf{Division of labour and the evolution of multicellularity}}
\author{Iaroslav Ispolatov$^1$, Martin Ackermann$^2$ \& Michael Doebeli$^3$  \\\\
\vspace{-2mm}\normalsize $^{1,3}$Departments of Zoology and Mathematics\\
\vspace{-2mm}\normalsize University of British Columbia, 6270 University Boulevard\\
\vspace{-2mm}\normalsize Vancouver B.C. Canada, V6T 1Z4\\
\vspace{-2mm}\normalsize $^{2} $Department of Environmental Sciences, ETH Zurich, and\\
\vspace{-2mm}\normalsize Department of Environmental Microbiology, Eawag\\
\vspace{-2mm}\normalsize $^1$Departamento de F\'isica\\
\vspace{-2mm}\normalsize Universidad de Santiago de Chile\\
\vspace{10mm}\normalsize Casilla 307, Santiago 2, Chile\\
\vspace{-2mm}\normalsize $^{1,3}$ To whom correspondence may be addressed. \\
\vspace{-2mm}\normalsize E-mail: doebeli@zoology.ubc.ca, jaros007@gmail.com}
\date{\normalsize\today}

\begin{document}


\def\T{\Theta}
\def\D{\Delta}
\def\d{\delta}
\def\r{\rho}
\def\p{\pi}
\def\a{\alpha}
\def\g{\gamma}
\def\ra{\rightarrow}
\def\s{\sigma}
\def\b{\beta}
\def\e{\epsilon}
\def\G{\Gamma}
\def\om{\omega}
\def\pe{$1/r^\a$ }
\def\l{\lambda}
\def\f{\phi}
\def\w{\psi}
\def\m{\mu}
\def\t{\tau}
\def\c{\chi}

\maketitle

\newpage
\begin{center} {\bf \large Abstract}
\end{center}
\noindent
Understanding the emergence and evolution of multicellularity and cellular differentiation is a core problem in biology.
We develop a quantitative model that shows that a multicellular form emerges from genetically identical unicellular ancestors when the compartmentalization of poorly compatible physiological processes into component cells of an aggregate produces a fitness advantage. This division of labour between the cells in the aggregate occurs spontaneously at the regulatory level due to mechanisms present in unicellular ancestors and does not require any genetic pre-disposition for a particular role in the aggregate or any orchestrated cooperative behaviour of aggregate cells.  Mathematically, aggregation implies an increase in the dimensionality of phenotype space that generates a fitness landscape with new fitness maxima, and in which the unicellular states of optimized metabolism become fitness saddle points. Evolution of multicellularity is modeled as evolution of a hereditary parameter, the propensity of cells to stick together, which determines the fraction of time a cell spends in the aggregate form.  Stickiness can increase evolutionarily due to the fitness advantage generated by the division of labour between cells in an aggregate.

\newpage
\vskip 2cm 

\section  {Introduction} 

Life on Earth takes a myriad of different forms, and understanding the evolution of this complexity is one of the core problems in all of science. The origin of species and the diversity of ecosystems are paradigmatic representatives of evolving complexity, but similarly fundamental questions arise when studying the evolution of multicellularity and cell differentiation. The evolutionary transition from unicellular to multicellular organisms is often referred to as one of the major transition in evolution (\cite{maynardsmith_szathmary1995}), even though many of the requirements for multicelluarity probably evolved in unicellular ancestors, thus facilitating the transition (\cite{Grosberg:2007}). The evolution of multicelluarity is characterized by the integration of lower level units into higher level entities, and hence is associated with a transition in individuality (\cite{buss1987, michod2007, michod1996, michod1997}). Such transitions are thought to be based on cooperation between the lower level units (\cite{buss1987, maynardsmith_szathmary1995, michod2007}), and recent models for the evolution of multicellularity are based on the concept of division of labour (\cite{gavrilets2010, willensdorfer2009, rossetti_etal2010}), typically between some and germ cells ({\cite{michod2007}). However, the existing models and explanations for the emergence of multicellularity provide only partial answers and raise further questions. In most models, some basic and pre-existing differentiation is assumed, and the circumstances under which such differentiation can be enhanced and stably maintained are investigated. To study the evolution of cell differentiation it is also often assumed that undifferentiated cells already occur in multicellular aggregates (\cite{gavrilets2010}), facilitated by selection on size due to environmental pressures (\cite{bonner1998}), such as predation (\cite{bell1985, boraas_etal1998}}) or the need for cooperation (\cite{pfeiffer_bonhoeffer2003}).

In this paper we consider the simultaneous evolution of multicellularity and cell differentiation in a population of identical and undifferentiated unicells, based on   the idea that the emergence of multicellularity and subsequent cellular specialization are driven by the fitness advantages of a division of labour between cells. Such a division of labour need not necessarily occur in the form of soma and germ cells. 
Even simple, unicellular organisms need to perform physiological tasks that cannot be efficiently accomplished simultaneously by the same cell. Examples include biochemical incompatibility between metabolic processes, such as between oxygenic photosynthesis and oxygen-sensitive nitrogen
fixation in cyanobacteria (\cite{flores2009,fay1992,berman2003, rossetti_etal2010}); motility and mitosis, processes which compete for the use of  the same cellular machinery, the microtubule organizing center (\cite{king2004}), and in general, reproduction and survival in challenging environment (\cite{nedelcu2006,kirk2003, Rainey:2010bh}).
Many unicellular organisms have overcome this problem by temporal segregation of incompatible activities, essentially cycling between phases dedicated solely to a single activity.  These cycles can be regulated by endogenous rhythmic mechanism as well as by external signals (\cite{fay1992,berman2003}).
Other cells found alternative means of limiting the detrimental effects of such incompatibility, such as introducing intracellular segregation, or limiting one activity to the minimum necessary for survival, or producing additional substances that chemically prevent the harmful interactions.

In a multicellular organism, such incompatible processes can take place simultaneously, but compartmentalized to separate cells. A first well-studied example of emerging intercellular separation of poorly compatible activities is the germ-soma specialization in volvox (\cite{nedelcu2006,kirk2003}). Somatic cells gather nutrients from the environment, and provide germ cells with these nutrients (\cite{bell1985, KOUFOPANOU:1994hq, Solari:2006kh}). The somatic cells are flagellated, and the flagella are important for motility and transport of nutrients to the cells (\cite{Solari:2006kh, Short:2006pt}). Flagellation and cell division are incompatible, and this fact is probably one of the factors promoting differentiation between somatic and germ cells (\cite{KOUFOPANOU:1994hq}).

A second well-studied example is the incompatibility between photosynthesis and nitrogen fixation, and the resolution of this incompatibility in filamentous cyanobacteria (\cite{flores2009,fay1992,berman2003, rossetti_etal2010}). The key enzyme for nitrogen fixation, nitrogenase, is sensitive to oxygen, and is thus inhibited by oxygenic photosynthesis (\cite{fay1992}). In filamentous cyanobacteria, this conflict is resolved by a spatial segregation of the two processes. A small proportion of the cells differentiate into heterocysts that fix nitrogen and do not engage in photosynthesis (\cite{Yoon:2001dw}), and these heterocysts exchange substrates with the vegetative cells in the same filament. 

The fitness advantage of such a division of labour is an important, if not crucial, factor in the emergence and evolution of multicellularity and cell specialization. In fact, the unicellular ancestors often already posses the prototypes of regulatory mechanisms that are needed to maintain cell specialization in the multicellular forms.  Consider an example of two incompatible processes $A$ and $B$ that are alternating in time in a single-cell organism and assume that the cell has developed a regulatory mechanisms that allows it to suppress the  process $A$ when the contrasting process $B$ occurs.  When two or more such cells come into a sufficiently close and long enough contact that allows them to exchange the benefits produced by these two processes, it may become more beneficial to end cycling in each cell and come to a steady state with one cell specialized in $A$ and the other in $B$. At the basic cellular signaling level, the endogenous mechanism that drives unicellular cycling is often based on accumulation of the products of $A$ or $B$ during the active phase and subsequent depletion during the passive phase
(\cite{nedelcu2006,kirk2003, berman2003}). Hence, when a partner cell keeps producing the product  $A$, the cell which produces $B$ does not experience the shortage of the product of $A$, which may bring the phase-changing mechanism to a halt. 
This principle can equally be applied to germ-soma specialization where $A$ and $B$ can be interpreted as reproduction and motility: reproductive cells may not run out of nutrients if they are repositioned by the soma cells of the colony to new feeding positions, while the soma cells do not die out as the reproductive cells keep on producing their genetically identical copies. 

The assumption that incompatible cellular processes suppress each other is supported by empirical evidence. Experimental work with Volvox carteri has shown that cells that are destined to become reproductive suppress the expression of genes encoding somatic functions, and somatic cells suppress germ cell functions (\cite{TAM:1991fc, Meissner:1999ay}). Also, gene regulatory mechanisms in unicellular ancestors can readily be co-opted during the transition to multicellularity, and contribute to differential gene expression in somatic and germ cells (\cite{nedelcu2006}).

Even if a unicellular form evolved to combine incompatible processes via other mechanisms, e.g. through the production of costly means of mediation of harmful interactions between metabolites, cells can often benefit from the opportunities that emerge in the multicellular form as well, e.g. when 
compartmentalization of 
contrasting processes in separate cells allows them to stop producing the costly mediation metabolites.
It is important to note that when single cells merge to form a two- or multi-cellular aggregate with permanently specialized cellular functions, no initial genetic specialization is required. Merging cells can be completely genetically identical, yet the symmetry of the initially unspecialized aggregated cells is broken spontaneously by regulatory mechanisms that function independently in each cell.  

Mathematically, the evolutionary advantage of the division of labour in aggregate forms can be viewed as the emergence of new, higher fitness maxima when the dimensionality of phenotype space is increased. The new fitness maxima are not a direct consequence of aggregation, but are based on the interaction between aggregated individuals that engage in the division of labor. An increase in the dimensionality of phenotype space occurs when two or more cells couple their metabolism by exchanging metabolites. Because of this exchange, the fitness of an individual cell depends not only on the metabolic state of itself, but also on the metabolic states of its aggregation partners, and a maximum in low-dimensional space describing the physiology of a single cell may often become a saddle point (i.e., a point that is a fitness maximum in one direction in phenotype space, and a fitness minimum in another direction) when new dimensions, i.e., new cells are added. However, the physiological state of each cell, 
such as the activity level of different processes, is regulated only by the cell itself based on the external and internal cues.  Thus, the new fitness maxima have to be achievable by independent regulatory adjustments of each cell.

In this paper we develop a model that serves as a quantitative description of the qualitative discussion above. 	
For the case of two incompatible processes, we will show that a sufficient condition for reaching a higher fitness state of cellular specialization in aggregates is the existence of unicellular regulatory mechanisms that suppress one process when the other is active, and vice versa. More specifically, we derive conditions in terms of the fitness function that favour the existence of a saddle point, and hence the evolution of division of labour, in 2-cell aggregates. We also show that the fact that higher fitness maxima can be attained in aggregate forms in turn selects for adhesion mechanisms that allow cells to form aggregates in the first place. We illustrate the general arguments by choosing particular forms of fitness costs and benefits and consider a simplified scenario of exchange of metabolites, and evolution of adhesion. Thus,
the results show that the emergence of multicellularity and cell differentiation can, in theory, result from the evolution of the propensity of cells to aggregate driven by the fitness advantage of division of labour in the aggregate forms. This advantage is due to incompatible physiological processes being performed  separately  in dedicated specialized cells, such that  the results of these processes are shared among the cells in the aggregate. 
The division of labour occurs spontaneously at the regulatory level due to mechanisms already present in unicellular ancestors and does not require any genetic pre-disposition for a particular role in the aggregate. 
 
\section{Model definition}

We envisage  the simplest possible scenario of aggregation,  the formation of a union of two cells, and we consider a population of cells that reproduce, die, and between birth and death can exist in uni-cellular or two-cellular forms. The population densities of single cells and two-cellular aggregates are denoted by  $n_1$ and $n_2$. We also assume that the transition between uni-cellular to two-cellular forms are reversible and may occur a number of  times during a cell's lifespan.  The binding  constant, which determines the fraction of time a cell spends in the two-cell form, is controlled by a heritable (genetic) parameter  $0 \leq\s \leq 1$, which we call the cell stickiness.
The total rate of aggregation $A_{i,j}$ between cells of stickiness types $\s_i$ and $\s_j$  is then given by     

\begin{equation}
\label{a}
A_{ij}= k_{+} n_1(\s_i)\s_i n_1(\s_j)\s_j,
\end{equation}
where $k_{+}$ is an aggregation constant, which we assume to be  identical for all cells, and $n_1(\s_l)$ is the population density of single cells with stickiness $\s_l$. 
Also, for simplicity we
 assume that rate of dissociation of an aggregate is independent on the stickiness of its (two) constituents. The per capita dissociation rate is denoted as $k_{-}$.  Together, these assumptions make the fraction of time a cell spends in the aggregate state an increasing function of its stickiness.  If a cell in an aggregate dies, the remaining cell becomes a single cell. If a cell in an aggregate divides, the daughter cell is released as a free cell while the aggregate remains intact. 

To measure ``fitness'', we assume that cells can produce two metabolites $x$ and $y$, and that production of these metabolites confers a benefit $B(x,y)$ and has a cost $C(x,y)$. The rate of reproduction of a cell that has metabolic rates $(x,y)$, i.e., its fitness, is then determined by the functions $B$ and $C$.  
For the single cell state, a simple form of the benefit function reflects the requirement that both metabolites are essential for the normal functioning of the cell:

\begin{equation}
\label{b}
B(x,y)=xy
\end{equation}

Any realistic cost function $C$ should satisfy the following constrains: 
First, since metabolic rates cannot increase indefinitely,
the fitness must have a maximum or maxima at some intermediate metabolic rates and rapidly decrease for high metabolic rates.  Hence the cost function $C$ should grow faster than $B$ in any direction in  the $x-y$ plane. Second, to incorporate our assumption that the production of $x$ and $y$ are poorly compatible cellular processes, the cost function $C$ should exhibit an ``inefficiency penalty'' for producing both metabolites $x$ and $y$ in the same cell. To incorporate these assumptions we consider cost functions of the form 

\begin{equation}
\label{c}
C(x,y)=c_x x^3 \exp(y^2) + c_y y^3 \exp(x^2).
\end{equation}

Essentially, this form means that if only one metabolite is produced (for example, $y=0$), the cost grows fairly slowly (algebraically),  $\sim x^3$, but when both metabolites are produced, $x=y$, the cost grows much faster, $\sim x^3 \exp(x^2)$. There are many forms of cost functions (e.g. algebraic ones,  $C(x,y)=c_x x^3(1+y^4) + c_y y^3(1+x^4)$) that satisfy these two constraints and would lead to very similar results as presented below.   

We note that this equation captures aspects of an empirically measured cost function (\cite{Dekel:2005yo}), namely that the metabolic costs associated with expressing a trait increase faster than linear with the expression level. The ``inefficiency penalty'' does not have a direct empirical basis, but represents the general principle of poorly compatible cellular processes. The importance of this will be explained in more detail below.

Our choice of $B$ and $C$ leaves only 2 free parameters,  $c_x$ and $c_y$, and to facilitate the visualization of the fitness landscapes (see below), we assume symmetry, i.e., $c_x=c_y\equiv c$. This assumption is not crucial for the results reported.

Metabolites produced by each cell are consumed by the cell itself when it is in the unicellular form,
or are assumed to be equally shared for consumption within a two-cell form. Hence the two-cell benefit function takes the form 

$$
B_{1,2}= \left(\frac{x_1+x_2}{2}\right) \left(\frac {y_1+ y_2}{2}\right).
$$
At the same time, each cell bears the costs of everything it produces in either single- or multi-cell state. 
Overall, fitness, i.e., the cellular rate of reproduction, is given by the difference between the benefits and the costs of metabolism. For a a single cell with metabolic rates $(x,y)$, fitness is therefore  

\begin{equation}
\label{r1}
R_1(x,y) = xy - [c x^3 \exp(y^2) + c y^3 \exp(x^2)].
\end{equation}
On the other hand, the individual fitness of a cell $i$ $(i=1,2)$ in a two cell aggregate with metabolic rates $\{x_1 ,  x_2,  y_1,  y_2\}$ is

\begin{equation}
\label{r2}
R_{2,i}(x_1,x_2, y_1, y_2) =  \frac{(x_1+x_2)(y_1+ y_2)}{4} - [c x_i^3 \exp(y_i^2) + c y_i^3 \exp(x_i^2)].
\end{equation}

Note that $R_1(x,y)=R_{2,i}(x,y,x,y)$, i.e. the fitness of a unicellular form is equal to the fitness of
each cell in the two-cellular form when both cells are producing the same amount of metabolites $x$ and $y$ (when analyzing the model dynamics below, we assume a cost of ÔstickinessÕ. Cells in two-cell aggregates tend to have higher stickiness, and thus pay an additional cost). 

Due to the symmetry of $R_1$, and because of the cost of producing high levels of metabolites, the fitness of a single cell will be maximized at some intermediate level of metabolite production $(x,y)=(x^*,x^*)$. Under certain conditions, this single-cell maximum becomes a saddle point for the fitness function of cells in a 2-cell aggregate. These conditions can be most easily seen by assuming complete symmetry between the two cells of a 2-cell aggregate, i.e., by setting $x_1=y_2$ and $x_2=y_1$ in (\ref{r2}). The fitness function of a cell in a 2-cell aggregate, which a priori is a function of four variables, then becomes 

\begin{align}
\label{Fit}
F(x,y)= R_{2,i}(x,y, y, x)
\end{align}
($i=1,2$), i.e., a function of two variables. If the function $F$ is restricted to the diagonal $x=y$, the single cell maximum $(x^*,x^*)$ can be recovered as a maximum along this diagonal, and the question is under what conditions this point becomes a saddle point, i.e., a minimum along the anti-diagonal.

In the Appendix it is shown that the main criterion for $(x^*,x^*)$ to be a saddle point is 

\begin{align}
\label{DF2}
\frac{\partial^2F}{\partial x\partial y}(x^*,x^*)<0.
\end{align}

This essentially means that the production of the two metabolites should be anti-synergistic in the vicinity of $(x^*,x^*)$, so that the gain from producing more of both metabolites is less than linear. For example, this can occur with the inefficiency penalty assumed for the cost function (\ref{c}). If this anti-synergy is strong enough, the single cell optimum $(x^*,x^*)$ becomes a saddle point for the fitness function of cells in 2-cell aggregates. Indeed, by construction of (\ref{r1},\ref{r2}), there exists a range of parameter $c$ for which the maximum
(in $\{x_1,x_2,y_1,y_2\}$ space) of the reproduction rate for a cell that is a part of a two-cellular complex is higher than that of a single cell. This is illustrated in Figure 1 for $c=1/25$. 

\begin{figure}
\includegraphics[width=.4\textwidth]{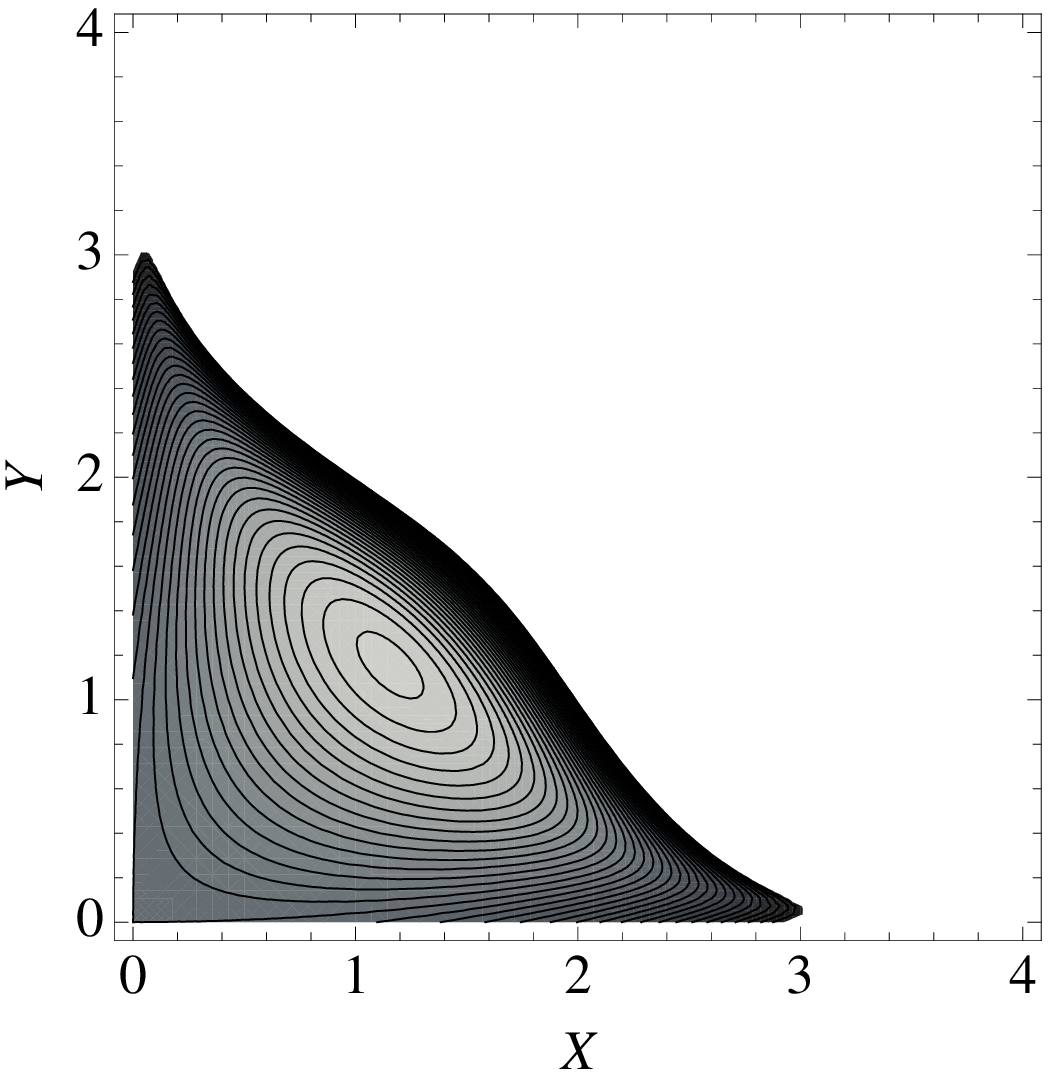}
\includegraphics[width=.4\textwidth]{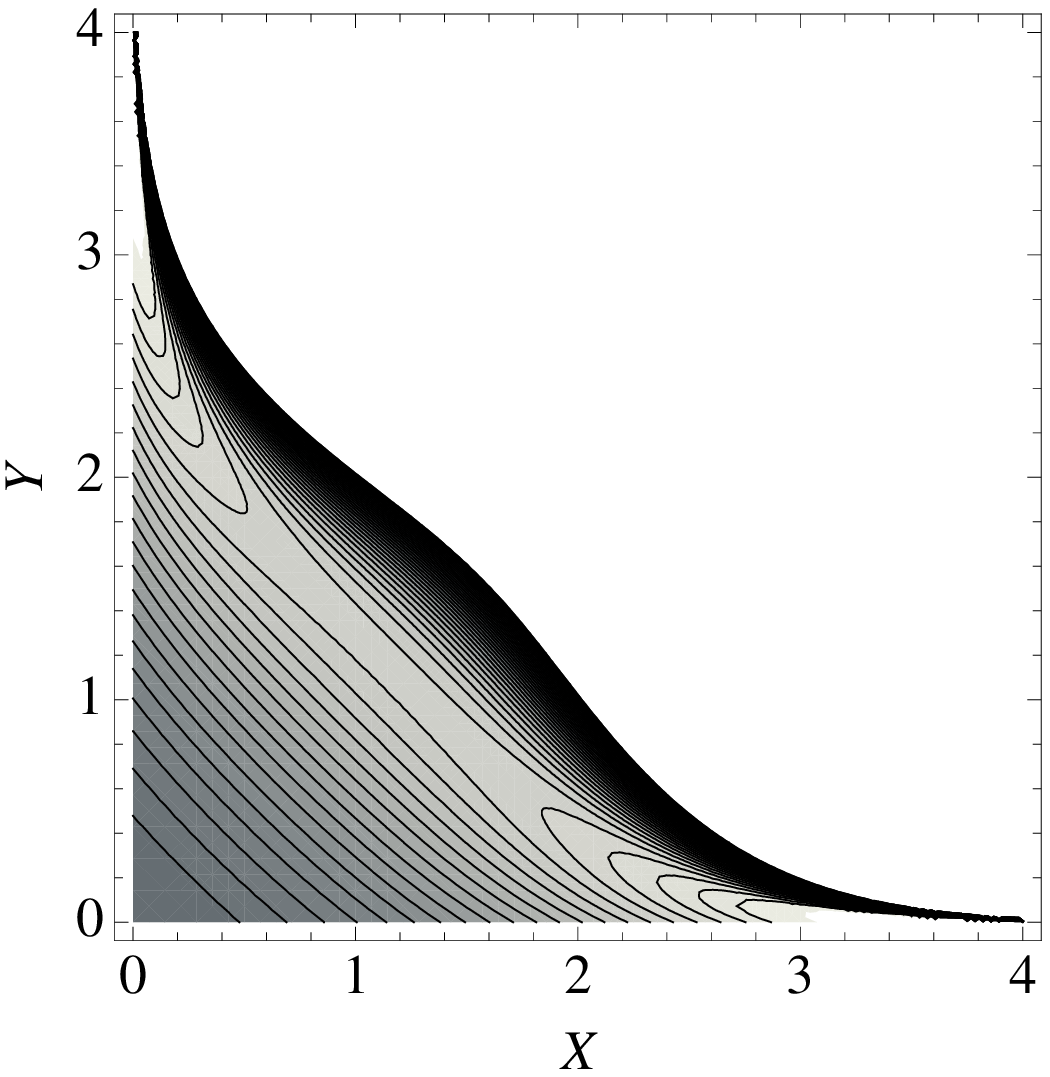}
 \footnotesize  
 \caption{\label{f1} 
The metabolic fitness landscape for single cells (defined by Eq.~(\ref{r1}), left panel) and for the corresponding two-cell aggregates (defined by Eq.~(\ref{r2}), right panel). For the two-cellular fitness we have assumed an anti-symmetry between the metabolic states of the cells, $x_1=y_2$ and $x_2=y_1$. The form of the fitness landscape (\ref{r2}) implies that the maxima visible in the right panel remain the same in the unrestricted 4-dimensional space $\{x_1,y_1,x_2,y_2\}$. The figure shows  how a maximum on the diagonal $x=y$ of the left panel, corresponding to equal rates of production of $x$ and $y$, becomes a saddle point for the two-cell fitness landscape in the right panel. Two maxima near the horizontal and vertical axes in the right panel correspond to compartmentalization of production of $x$ and $y$ in the two-cell state: while one cell produces only $x$, the other cell produces only $y$. The white area in the plots corresponds to negative birth rates (i.e., $B-C<0$). The landscapes shown correspond to $c_x=c_y=c=1/25$.
}
\end{figure}

The left panel depicts the single-cell state fitness $ R_1(x,y)$, which has a maximum at the diagonal $x=y=x^*$ with $R_1^{max}\approx 0.866$. The right panel shows the fitness of a cell in a two-cell aggregate. With the assumption that we made above on the  $x-y$ symmetry of the cost function, the maxima of the two-cell-state fitness are in the  subspace when two cells are exactly in anti-symmetric metabolic states, $x_1=y_2\equiv x$, and $x_2=y_1\equiv y$, hence they are maxima of the fitness function (\ref{Fit}) (here $F^{max}=625/432\approx 1.45$). One can compare the two panels by noting that the diagonal $x=y$ in both of them corresponds to the same function $F(x,x)$. But while for the single-cell state the two-variable fitness $R_1(x,y)$ has a global maximum along the diagonal $x=y=x^*$ (left panel), the two-cell state fitness $F(x,y)$  has a saddle point at $x=y=x^*$ on the diagonal (right panel). For larger $c$ this condition is no longer satisfied, so the two-cell state has the same diagonal fitness maximum as the single-cell form, and division of labour would not be expected.

It should be noted that condition (\ref{DF2}) is a local condition near the single cell maximum $(x^*,x^*)$, and that the cost functions (\ref{c}) is of course only one of many such functions that lead to fitness functions that can satisfy this condition (\ref{DF2}). Thus, the cost (and benefit) function used here merely serve to illustrate a general principle. 

To complete the basic model description, we assume that cells, whether in uni- or two-cellular form, reproduce individually by  periodically releasing unicellular offsprings at a rate that is proportional to the fitness of a cell.  The only cell property that is inherited during reproduction is the stickiness $\s$. A small random variation in $\s$  occurring during reproduction corresponds to mutations in this genetically determined trait.  Finally,  we assume a logistic form of the per cell death rate $Dp$, which is independent of whether a cell is single or a part of an aggregate,

\begin{equation}
\label{d}
Dp=\d N, \;\; N=\int_{\s} [n_1(\s) + n_2(\s)] d\s;
\end {equation}
where $\d$ is a parameter and $N$ is the total population.

\section{Model dynamics}
There are three biologically distinct time scales in our model:  the fast  regulatory metabolic adjustment, the intermediate rate of cellular aggregation and dissociation,  and slow reproduction and concomitant evolution of the heritable trait $\s$.  This natural time scale separation allows us to simplify the mathematical analysis, assuming steady states of the faster processes in the dynamics of the slower events. 

\subsection{Metabolic regulation}
So far we have specified benefits and costs of producing two metabolites at rates $x$ and $y$, but we have not specified what controls the dynamics of metabolic regulation of $x$ and $y$. 
The basic assumption is that each cell adjusts the rates of production of the metabolites to maximize fitness for given conditions (unicellular or aggregate form) via a fast regulatory mechanism.  The important part of this assumption is that each cell acts individually to maximize its fitness, without ``coordinating''  the metabolic regulation with its partner, yet the conditions in which a cell operates do depend on whether or not the cell is in aggregate form, and if it is, on the metabolic state of the partner. We first assume that in the single cell state, each cell has a naturally occurring mechanism for regulating its metabolite production to the optimum of the fitness landscape given by $B(x,y)-C(x,y)$, with $B$ and $C$ the cost functions \ref{b} and \ref{c}. The  fitness landscape in metabolic space defined by Eq.~(\ref{r1}) and illustrated in Figure 1 has a fairly simple form, so it seems reasonable to assume the regulatory convergence to the unicellular fitness maximum.

For example, one can assume that the cells adjust the metabolic states following the gradient in the fitness landscape with some random noise $\eta$,

\begin{align}
\label{stoch1}
\nonumber
\frac{d z_i}{dt} = \a \frac {\partial R}{\partial z_i} + \eta_{zi}(t)\\ 
\langle \eta_{zi}(t)\rangle = 0, \;\; \langle \eta_{zi}(t) \eta_{zj}(t')\rangle =
\delta_{i,j} \delta (t-t') \Gamma
\end{align}
Here $z_i$ is one of the two metabolic coordinates $\{x,y\}$,
$\eta_{zi}(t)$ is a random noise term with zero mean that is assumed uncorrelated in time and between different coordinates. 
A process given by (\ref{stoch1}) converges to a steady state distribution of the cell population in ``metabolic space'' $n(z)$ (\cite{vanKampen}),

\begin{equation}
\label{steady}
n(z) = n_0 \exp \left[ \frac {2\a}{\Gamma} R(z)\right],
\end{equation}
where $z = (x,y)$ 
are again the metabolic coordinates in the unicellular state. We assume that the dynamics is fast (large $\alpha$) and the noise is weak (small $\Gamma$), so that the population quickly becomes concentrated in the vicinity of the metabolic state conferring maximum fitness.

We assume that the same regulatory mechanisms that lead to metabolic fitness maximization in the unicellular state regulate the metabolic rates of two cells in aggregate form to a metabolic fitness maximum for the two-cell aggregate (essentially, the noise term in the metabolic dynamics (\ref{stoch1}) leads to Ôsymmetry breakingÕ, enabling two cells that just aggregated to diverge in their metabolic phenotypes). In fact, the dynamics given by (\ref{stoch1}), applied to $z_i$ being one of the four coordinates $\{x_1,y_1, x_2, y_2 \}$, also lead to a concentration of the metabolic rates $(x_1,y_1, x_2, y_2)$ of two aggregated cells in the vicinity of a fitness maximum. The corresponding fitness landscape is illustrated in Figure 1 (right panel). The Figure illustrates that the fitness maximum of a cell in a two-cell aggregate can be higher than the metabolic fitness maximum attained by a single cell. Essentially, this is because the fitness maximum of a single cell, which lies on the diagonal $x=y$, becomes a saddle point in the higher-dimensional metabolic fitness landscape of the aggregate form.

In the following, we therefore assume that each cell is at a metabolic state that maximizes its reproductive rate. This allows us to drop the metabolic coordinates from the notation for the cell concentration.  For example, for $c_x=c_y=1/25$, the (steady state) reproduction rate of a single cell becomes  $R_1\approx 0.866$ (maximum on the diagonal of the left panel of Fig.~1) and the reproduction rate of a cell in a two-cell state $R_2=625/432\approx 1.45$ (corresponding to the maxima in the right panel of Fig.~1)

\subsection{Transition from unicellular to multicellular states}
The kinetics of the association of two cells with stickiness $\s_1$ and $\s_2$ into a two-cell complex with concentration $n_c(\s_1,\s_2)$ is described by the following rate equation,
\begin{equation}
\label{ass}
\frac{\partial n_c(\s_1,\s_2)}{\partial t }=  k_{+} n_1(\s_1) \s_1 n_1(\s_2) \s_2
- k_{-}n_c(\s_1,\s_2) - 2 \d N n_c(\s_1,\s_2).
\end{equation}
Here the gain or association term has the form introduced in (\ref{a}),
the first loss term describes the dissociation of a two-cell complex into individual cells and the second loss term describes the loss of a two-cell complex due to a death of one of its constituents as defined in (\ref{d}). 
The factor 2 reflects the fact that a death of any of two cells is sufficient for the elimination of a two-cell aggregate.
Assuming sufficiently fast association and dissociation compared to the time scale of the evolution of stickiness, we calculate the steady state concentrations of two-cell complexes for a given density of  single cells,
\begin{equation}
n_c^*(\s_1,\s_2)= \frac {k_+ n_1(\s_1) \s_1 n_1(\s_2) \s_2} {k_- + 2 \d N}.
\end{equation}

To determine the steady state concentration $n_2(\s)$ of individual cells in aggregates,
the concentration of complexes $ n_c^*(\s_1,\s_2)$ has to be integrated over one of its coordinates,

\begin{align}
\label{stst}
n_2^*(\s)=2\int_0^{\infty} n_c^*(\s,\s') d\s'= &\ n_1(\s) \chi\\ 
\nonumber
\chi = &\ \frac {2k_+\s \int_0^{\infty} n_1(\s') \s' d\s'} {k_- + 2 \d N}.
\end{align}
The factor 2 reflects the fact that either of two cells in an aggregate can have the stickiness $\s$. 
Note that $\chi$  needs to be determined self-consistently since the quantity $N$ in the denominator depends on the total number of cells, $N=N_1+N_2$, $N_j=\int n_j(\s) d\s$.

A rate equation that is analogous to (\ref{ass}) holds for the evolution of the single-cell population:

 \begin{align}
\label{diss}
\nonumber
\frac{\partial n_1(\s)}{\partial t }=& - 2 k_{+} n_1(\s)\s \int_0^\infty  n_1(\s')\s'd\s'
+ k_- n_2(\s)  + \d N n_2(\s) - \d N n_1(\s)\\
&+ \left[ R_2 n_2(\s) + R_1 n_1(\s) \right] M(\s) 
\end{align}
Here the first loss term denotes the association of a single cell $\s$ into an aggregate with any other single cell (with the factor 2 describing that two single cells are lost), the first gain term describes the dissociation of an aggregate, the second gain term describes the appearance of a single cell when a cell in an aggregate dies, and the second loss term corresponds to the loss of a single cell due to its death. The gain term in the third line accounts for the appearance of a new single cell due to cell division, which occurs within  aggregated and free cells with rates $R_2$ and $R_1$, defined by eqs. (\ref{r1}) and (\ref{r2}). The term $M(\s)$, a decreasing function of $\s$, accounts for the fitness costs of maintaining stickiness. This term imposes a penalty for two-cell aggregates; aggregation can thus only evolve if it provides a non-negligible advantage to the aggregating cells.

\subsection{Evolution of stickiness}
Substituting (\ref{stst}) into (\ref{diss}) we arrive at the equation for the evolution of the  population density of cells with stickiness $\s$,

\begin{align}
\label{evol}
\frac{\partial n_1(\s)}{\partial t }=&-\d N n_1(\s)(1+\chi)\\
\nonumber
 &+M(\s) (R_2 \chi + R_1)  \left[ n_1(\s) + D \frac{\partial^2 n_1(\s)}{\partial \s^2}\right]. 
\end{align}
The diffusion term, proportional to a small constant $D$ and added to the birth term, describes mutational variation in stickiness at birth. Eq. (\ref{evol}) can be solved numerically. For suitable parameter combinations, the population evolves towards higher stickiness, resulting in cells spending most of their life in two-cell aggregates. 
\begin{figure}
\includegraphics[width=.4\textwidth]{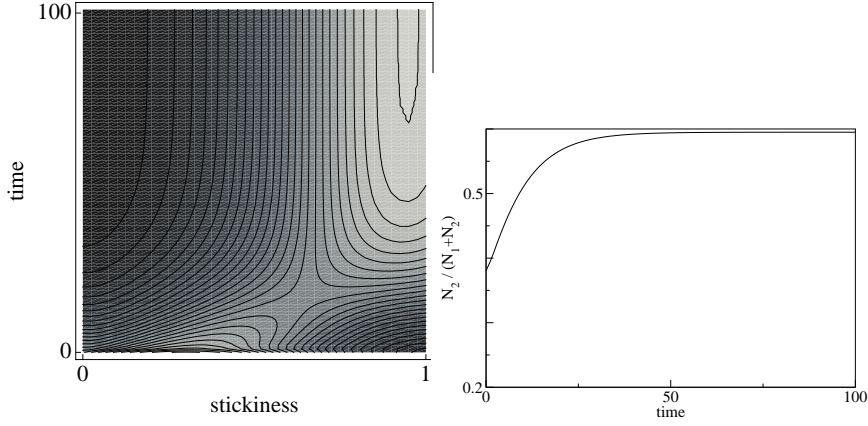}
 \includegraphics[width=.4\textwidth]{f2b.eps}
 \footnotesize  
 \caption{\label{f2} 
Evolution towards cell aggregation and two-cellularity. Panel a) shows the population distribution of the trait $\s$ over time, with brighter areas indicating higher densities. Panel b) shows the proportion of 2-cell aggregates, $N_2/(N_1+N_2)$ with $N_j=\int n_j(\s)d\s$ for $j=1,2$ as a function of time. For the figure, eq.~(\ref{evol}) was solved numerically for the following parameters: 
$k_+=10$, $k_-=1$, $\d=1$, $M(x)=1-x/10$, and $D=10^{-3}$. As mentioned earlier, the effective birth rates for unicellular and two-cellular forms were assumed to be perfectly optimized to the corresponding cellular state,  $R_1 = 0.866$ and $R_2 = 1.45$.
}
\end{figure}
This process is illustrated in Figure 2 and is due to the higher birth rates and resulting evolutionary advantage of cells that are more sticky and thus spend more time in a state that is metabolically superior due to division of labour. 

\section{Discussion}
The evolution of cell differentiation in multicellular aggregates is an important transition in the history of life on Earth. Most existing models of this transition assume some pre-existing differentiation in the single cells and/or the pre-existence of some form of compartmentalization, i.e., multicellularity (e.g. \cite{rossetti_etal2010, willensdorfer2009, gavrilets2010}). The main result of the present study is a proof of principle that multicellularity and cell specialization can emerge in genetically and physiologically homogenous populations via spontaneous breaking of cellular universality (or symmetry) by regulatory non-hereditary metabolic mechanisms. 

Such symmetry breaking occurs if the division of labour between cells brings certain fitness advantages and if regulatory mechanisms that allow cells to optimize their physiology exist in the ancestral unicellular form. It is important to note that with such mechanisms, the cells adjust their regulatory state individually, in a ``selfish'' way, thus no assumption about special cooperative interactions between the cells in an aggregate are necessary. Essentially, a suitable fitness landscape, exhibiting higher fitness for cells with differentiated functions in the aggregate form, determines the path of regulatory optimization towards cellular specialization. The prerequisite for this is that the 4-dimensional fitness landscape of a 2-cell aggregate has higher peaks than the 2-dimensional fitness landscape of a single cell, reflecting the advantages of division of labour. In other words, the maximum of the 2-dimensional landscape turns into a saddle point in the 4-dimensional landscape.  This qualitatively defines the 
general properties of the fitness function that promote (or inhibit) the transition to multicellularity, and can be formalized by considering second derivatives of fitness functions (see Appendix), which reveals that the main criterion for the maximum of the 2-dimensional landscape to turn into a saddle point in the 4-dimensional landscape is that there should be anti-synergistic interaction between the two metabolites, so that an increase  in the production of both metabolites in a single cell has a sufficiently high ÒinefficiencyÓ penalty. The cost functions used in this paper provide one example of such an inefficiency penalty. 

As a consequence of metabolic inefficiency, an effective increase in the dimensionality of the physiological pathways due to aggregation enables the cells to attain higher fitness in the aggregate form by dividing the labour of producing the two metabolites. If stickiness, i.e., the tendency to form aggregates, is a trait under selection, and if the costs of stickiness is not too large, then the increase in dimensionality of the fitness landscape and the concomitant increase in physiological fitness leads to the evolution of more sticky cells,   resulting in the emergence of multicellularity and cellular specialization within the aggregates. We note that this mechanism for the evolution of aggregation of single cells into multicellular clusters is different from the classic hypothesis that such aggregation is driven by some form of selection for size (\cite{bell1985,bonner1998}), e.g. due to predation (\cite{boraas_etal1998}) or the need for cooperation (\cite{pfeiffer_bonhoeffer2003}).

In virtually all models for cell differentiation, the basic underlying mechanism is a tradeoff between different physiological functions. It is then usually assumed that, for unspecified reasons, there exist different cell types that either already occupy different locations on the tradeoff curve (e.g. \cite{rossetti_etal2010, willensdorfer2009}), or have the genetic potential to do so (e.g. \cite{gavrilets2010}). Subsequently, the optimal composition of the different cell types in cell aggregates is studied. In contrast, in our model, all cells are in principle physiologically identical, and the differentiation only manifests itself through a purely physiological regulatory mechanism once cells occur in aggregates. In essence, the symmetry breaking regulatory mechanism generates a permanent spatial differentiation in cell aggregates. Such a physiological crystallization of potential temporal differentiation of single cells has been envisioned as one of the main routes to multicellular differentiation (\cite{nedelcu_michod2006}), and our model can serve as a basic metaphor for this process.

In most accounts, the basic physiological tradeoff underlying the transition to differentiated multicellularity is between reproduction and viability, and hence between soma and germ cells (\cite{gavrilets2010}, \cite{kirk2003}, \cite{michod2006,michod2007}). The model presented here can also be viewed in that context.
Then the physiological variables $x$ and $y$ become traits describing reproductive productivity (number of offspring) and viability (probability to survive to reproduction). 
The latter depends on many factors, such as the ability of cells to move, which in some types of organisms is incompatible with mitosis (\cite{king2004}). In a germ-soma specialization scenario, the regulatory mechanisms relevant for symmetry breaking may be based on a response to signals to stop growth that are emitted by fully developed bigger germ cells  (\cite{kirk2003}). Such signals could arrest the development of pro-soma cells, rendering them sterile. The predisposition for initial size and subsequent germ-soma differentiation would then stem from spontaneous asymmetric cell division (rather than from genetic differences). 

To critically evaluate the plausibility of our model for the evolution of multicellularity, it will be essential to test the main assumptions and predictions experimentally.  The most critical assumptions of the model are that i) some important cellular processes cannot be performed well in the same cell, ii) cells can readily evolve increased levels of attachment, and iii) attached cells can complement each other metabolically, and thus specialize on one of two poorly compatible processes. The first assumption, about trade-offs between cellular processes, if fundamental to most models of metabolic specialization (\cite{gudelj_etal2010}). A number of recent  studies have established concrete molecular mechanisms that can lead to trade-offs (\cite{carlson2007, knight_etal2006, molenaar_etal2009, scott_etal2010, beg_etal2007}), and it will be interesting to test whether such trade-offs are pervasive between different types of metabolic processes, and in many different organisms. If they are, this would increase the plausibility of the evolutionary transition towards multicellulary proposed here.

Testing the predictions of our model is challenging, but seems possible in principle. The main prediction is that conditions in which important cellular processes are incompatible with each other will promote the evolution of increased levels of attachment between complementary cell types. It is worthwhile considering whether this prediction can be tested with evolutionary experiments in the laboratory.  As discussed above, oxygenic photosynthesis and nitrogen fixation are prime examples of incompatible processes (\cite{bermanfrank_etal2003}), and unicellular cyanobacteria separate these processes temporarily, by performing photosynthesis during daytime and nitrogen fixation at night. One possible direct test of our model would be to evolve unicellular cyanobacteria in the laboratory under conditions where both processes are expected to be active i.e., in continuous light in medium without fixed nitrogen and ask whether the bacteria evolve adhesion and exchange of fixed compounds between cells that perform different processes. It is worth noting that the evolution of stickiness, i.e., of unicellular organisms forming multicellular clusters, has recently been observed in yeast (\cite{ratcliff_etal2011}), although the importance of incompatible metabolic process for this phenomenon remains to be determined.

In conclusion, in this paper we present a model showing that multicellularity and cellular differentiation can develop when cells can form an aggregate that enables them to exchange chemical signals and metabolites. This aggregate essentially has a higher physiological dimension, so that when there are cellular processes that are incompatible in a single cell, segregation of these processes into separate cells is possible in the aggregate form. Regulatory mechanisms that can control such a division of labour within an aggregate can be expected in many ancestral unicellular forms and are based on signals coming either from the cell itself, or from partner cells in the aggregate environment. The resulting division of labour can generate fitness benefits that lead to selection on the propensity of cells to aggregate, and hence to form multicellular and differentiated organisms. 

\vskip 1cm
\noindent {\bf Acknowledgements:} M.D. acknowledges the support of NSERC (Canada) and of the Human Frontier Science Program. I. I. acknowledges the support of FONDECYT (Chile). M.A. acknowledges the support of the Swiss National Science Foundation. 
\newpage 

\section{Appendix}

As explained in the text, the symmetry between $x$ and $y$, which follows from the form of the fitness functions (\ref{r2}), allows for a reduction of the dimension of metabolic space from four to two. Accordingly, we consider the symmetric fitness function $F(x,y)= R_{2,i}(x,y, y, x)$, $i=1,2$, where the $R_{2,i}$ are the fitness functions (\ref{r2}) of a single cell in a 2-cell aggregate. Then the restriction of $F$ to the diagonal $x=y$ is

\begin{align}
g(z)=F(z,z),
\end{align}

and the restriction of $F$ to the anti-diagonal through the point $(x^*,x^*)$ is
\begin{align}
h(z)=F(z,2x^*-z).
\end{align}

Along the diagonal, $(x^*,x^*)$ is a fitness maximum by assumption, hence

\begin{align}
\label{Dg2}
\frac{\partial^2g}{\partial z^2}(x^*,x^*)=\frac{\partial^2F}{\partial x^2}(x^*,x^*)+2\frac{\partial^2F}{\partial x\partial y}(x^*,x^*)+\frac{\partial^2F}{\partial y^2}(x^*,x^*)<0.
\end{align}

For $(x^*,x^*)$  to be a saddle point of $F$, it must be a fitness minimum along the anti-diagonal, hence we must have

\begin{align}
\label{Dh2}
\frac{\partial^2h}{\partial z^2}(x^*,x^*)=\frac{\partial^2F}{\partial x^2}(x^*,x^*)-2\frac{\partial^2F}{\partial x\partial y}(x^*,x^*)+\frac{\partial^2F}{\partial y^2}(x^*,x^*)>0.
\end{align}

It is clear that this last inequality inequality is satisfied if $\frac{\partial^2F}{\partial x\partial y}(x^*,x^*)$ is negative enough. We note that (\ref{Dh2}) also tends to be satisfied if the pure second derivatives of $F$ are positive at $(x^*,x^*)$, but this also tends to violate the condition (\ref{Dg2}) for $(x^*,x^*)$ to be a maximum along the diagonal. If symmetry between $x$ and $y$ is not assumed, similar considerations lead to analogous criteria in terms of second derivatives of fitness functions for a maximum in 2-dimensional to become a saddle point in 4-dimensional space.

\bibliography{multicell}
\bibliographystyle{prslb}

\end{document}